\documentclass{llncs}

\usepackage{amsmath,epsfig,epstopdf,amssymb}
\usepackage{adjustbox,lipsum}
\usepackage{url}
\usepackage{float}
\usepackage{xspace}
\usepackage{algorithmic}
\floatstyle{boxed}
\newfloat{algorithm}{htbp}{loa}
\floatname{algorithm}{Algorithm}
\floatstyle{boxed} 
\usepackage{longtable}
\usepackage{wrapfig}

\usepackage{multirow}

\newtheorem {observation} {Observation}

\floatstyle{boxed}
\newfloat{algorithm}{htbp}{loa}
\floatname{algorithm}{Algorithm}
\floatstyle{boxed} 
\usepackage{wrapfig}
\usepackage{multirow}
 \floatstyle{boxed} 
\restylefloat{figure}
\usepackage{wrapfig}
\newfloat{figure}{h}{ext}
\floatname{figure}{Figure}

\usepackage[font=small,labelfont=bf]{caption}
\DeclareCaptionFont{tiny}{\tiny}\usepackage{float}

\title{Termination of Monotone Programs}

\author  {Omar Al-Bataineh}
\institute {Nanyang Technological University } 
\date{}

\begin{document}
\label{firstpage}


\maketitle
\begin{abstract}
We present an efficient approach to prove termination of
monotone programs with integer variables, an expressive class of loops
that is  often encountered in computer programs. Our approach is
based on a lightweight static analysis method and takes advantage of simple 
properties of monotone functions.  
Our preliminary implementation 
shows that our tool has an advantage over existing
tools and can prove termination for a high percentage of loops
for a class of benchmarks.

\end{abstract}



\section{Introduction}

Proving termination of programs is a challenging and important problem
whose solutions can significantly improve software reliability and
programmer productivity.  Termination analysis also plays a role in
the analysis of reactive systems, non-terminating systems that engage
in ongoing interaction with their environments. In this case,
a termination argument is used to prove liveness properties such as the
absence of deadlock or livelock, by establishing that some desired
behaviour is not postponed forever. While the problem of proving
termination has now been extensively studied
\cite{Tiwari04,Podelski2004,Braverman06,Berdine2007,Gupta2008,Chawdhary2008,Gulwani09,Gulwani09speed,Kroening2010,Harris2010,TsitovichSWK11,Ben-Amram2012,LeikeH15},
search for efficient methods for proving non-termination that can cover a broad range of programs
remains open.

A huge body of work has been done on proving termination of programs
that is based on a number of techniques such as abstract
interpretation \cite{Berdine2007,Chawdhary2008,TsitovichSWK11}, bounds
analysis \cite{Gulwani09,Gulwani09speed}, ranking functions
\cite{Bradley2005,Cousot05}, recurrence sets
\cite{Gupta2008,Harris2010} and transition invariants
\cite{Kroening2010,Podelski2004}.  The most popular technique to prove
termination is through the synthesis of a ranking function, a mapping
from the state space to a well-founded domain, whose value
monotonically decreases as 
the computation progresses.
However, it is known that the linear ranking approach cannot completely resolve the
problem \cite{PodelskiR04,Braverman06}, since there are terminating
programs that have no such ranking function \cite{Ben-Amram2012}. Moreover,
linear ranking functions do not suffice for all loops, and, in
particular for multipath loops \cite{Ben-Amram2014}.  Thus, the
problem of decidability of termination for linear loops stays open, in
its different variants.

Monotonicity is a useful property that is often 
encountered in computer programs \cite{Gupta1990,Spezialetti1995}.  Indeed, a very
high percentage of variables of programs are monotonic in nature and
therefore the analysis of such programs can be significantly improved
by exploiting monotonicity.  The complexity of the termination problem
together with the observation that most loops in practice have
(relatively) simple termination argument and are monotone in nature
suggest the use of light-weight static analysis for this purpose.

The termination problem is a hard problem, and one can expect it to be decidable only in the simplest cases \cite{Braverman06}. In this work we study the termination problem of a simple class of loops with multiple paths and linear monotone assignments. The loop is specified by some initial condition, a loop guard, and a ``loop body'' of some restricted form.  
We allow the loop guard to be either a conjunction of diagonal-free constraints (i.e. where the only allowed comparisons are between a loop variable and a constant) or diagonal constraints (where difference between two variables can be also compared with constants). The analysis takes advantage of properties of monotone expressions \cite{Gupta1990,Spezialetti1995}, where we use recurrence relations to model the loop iteration update of each loop variable over a single variable $n \in \mathbb{N}$ denoting the loop counter. We consider a wide variety of loop forms with different syntactic structure (i.e. different forms of loop guard and update expressions). For each of these forms we identify exactly the concrete conditions under which the loop can be said to be non-terminating. These conditions turn out to have a surprising level of complexity. In particular, we find that these conditions can differ, depending on the assumptions that we make about the syntactic structure of the loop. 
An advantage of the proposed approach is that it works on programs as they are, without imposing any extra annotation effort or  invoking a safety checker as other methods typically do.

To demonstrate the efficiency of the proposed approach we compare our
implementation against the (strongest) tools that participate in
SV-COMP\footnote{\url{http://sv-comp.sosy-lab.org/2016/}} using two
benchmarks: the SNU real-time benchmark (\path{www.cprover.org/goto-cc/examples/snu.html}) and the Power-Stone benchmark
\cite{Ku2007}. Namely, we compare the implementation against the tools
AProVE \cite{AProveIJCAR14} and 2ls \cite{cdksw2015}.  Our implementation
outperforms both AProVE and 2ls in two aspects. First, it can handle a larger
class of monotone programs than both AProVE and 2ls. Second,
for the programs that all tools can handle our tool can decide
termination much faster.
 
\paragraph*{Related Work.}
Termination is a fundamental decision problem in program verification. In particular, termination of programs with linear assignments and guards has been extensively studied over the last decade. This has led to the development of powerful semi-algorithms to prove termination via synthesis of ranking functions, several of which have been implemented in software-verification tools such T2 \cite{Cook2006} and ARMC \cite{Podelski07armc}. The work in \cite{Colon2001,ColonS02,Bradley2005} use algorithms with sub-procedure for ranking individual paths; which is focused on the iterative construction of a termination argument for a full program. See the  surveys  by  Ben-Amram  and  Genaim \cite{Ben-Amram2014}, and by Gasarch \cite{Gasarch15a} describing semi-algorithmic approaches to termination based on ranking functions.

In \cite{Tiwari04} Tiwari proves that the termination of a class of single-path loops with linear guards and assignment is decidable when the domain of the variables is $\mathbb{R}$. Braverman proved that this holds for $\mathbb{Q}$ as well \cite{Braverman06}. However, both have considered universal termination of single-path loops, the termination of single/multi-path loops for a given input left open.  

Biere, Artho, and Schuppan propose an encoding of liveness properties into
an assertion \cite{Biere02livenesschecking}. This approach allows proving termination of programs without a ranking sub-procedure. It has been reported to prove termination of programs that require non-linear ranking functions. 
Prior experimental results on some benchmarks indicate this encoding results in difficult safety checks \cite{Cook2010}.

Berdine et al. present an algorithm for proving termination that is based on
abstract interpretation \cite{Berdine2007}. Using an invariance analysis they construct a variance analysis, 
and they use the fact that the transitive closure of a well-founded
relation is also well-founded to show that the fixed-point obtained by their analysis is correct.

\section{Preliminaries}

In this section we introduce some definitions that we use throughout the paper. 

\begin{definition}

A linear loop $\mathcal{L}$ with integer variables $x_1,..,x_n$ is a while loop of the form
$$
\textbf{while} ~ (\phi_1 \land ... \land \phi_{m}) ~ \textbf{do} ~\{s_1;...; s_n \}
$$
where each condition $\phi_i$ is one of the following form $(x_i \sim c)$ or $( (x_i - x_j) \sim c)$
such that $c \in \mathbb{Z}$ and $\sim \in \{<, \leq, >, \geq\}$. We call the constraints of the form $(x_i \sim c)$ as diagonal-free constraints and the constraints of the form $( (x_i - x_j) \sim c)$ as diagonal constraints. 
Each instruction $s_i$ is one of the following form
$$
s_i : = Assignment \mid \textbf{if} ~ \phi~ \textbf{then} ~ Assignment ~ \textbf{else} ~ Assignment
$$
where $\phi$ is a condition and $Assignment$ can be of the following form
$$
x_i := u \mid  x_i := u * x_i \mid x_i := x_i + v \mid x_i := u * x_i + v
$$
such that  $u, v \in \mathbb{Z}$. 

\end{definition} 
Intuitively, the semantics of a while loop can be interpreted as follows: starting from initial values for the variables $x_1,..., x_n$ (the input), the instructions $s_1,...,s_n$ are executed either sequentially or conditionally as far as the condition $(\phi_1 \land ... \land \phi_{m})$ holds. We say that the loop terminates for a given input if the condition $(\phi_1 \land ... \land \phi_{m})$ eventually evaluates to \textit{false}.

In this paper we are interested in studying termination of linear
monotone loops.  The property of monotonicity of a statement is
defined with respect to a specific loop surrounding the statement.
Consider a while loop $\mathcal{L}$ and a statement $\verb+s+: x :=
e;$ inside the loop. Further consider a single execution of the loop
which involves $n$ iterations through the loop. Let $\ell_1,
\ell_2,...,\ell_n$ denote the $n$ consecutive iterations of the loop
and $x_1, x_2,.., x_m$, denote the values assigned to $x$ during these
iterations, where $m \leq n$ because statement $\verb+s+$ may not be
executed during every iteration if there are conditional branches
inside the loop.

\begin{definition}[Monotonic statements \cite{Spezialetti1995}]
A statement $\verb+s+$ is considered to be loop monotonic w.r.t. loop
$\mathcal{L}$ if the sequence of values assigned to variable $x$
during successive executions of $\verb+s+$ forms an increasing or
decreasing sequence of values (i.e. $x_i < x_{i+1}$ or $x_i >
x_{i+1}$).
The monotonic statement $\verb+s+$ is considered to be regular
monotonic if the sequence $x_1, x_2,.., x_m$ is an arithmetic
progression or geometric progression. Otherwise, the monotonic
statement is considered to be irregular monotonic.
\end{definition}

The class of monotonicity of a statement $\verb+s+: x := e;$
surrounded by a loop $\mathcal{L}$ can be determined using a set of
static techniques. In \cite{Spezialetti1995} a sophisticated static
analysis is employed in order to determine loop monotonic
variables. For space reasons we omit discussion of these techniques
and we refer the reader to \cite{Spezialetti1995}. Throughout this
paper we use the following notations: $\uparrow$ to denote a
monotonically increasing update expression, $\downarrow$ to denote a
monotonically decreasing update expression, and $\rightarrow$ to denote a
constant expression.

\section{Single-path Linear Monotone Loops}

In this section we consider termination of single-path linear monotone loops. We consider first 
termination of single-path loops with diagonal-free constraints and then single-path loops with diagonal constraints.
The structure of single-path loops with diagonal-free constraints that we consider has the following form
\begin{equation} \label{cyclesWithoutCond}
 \begin{array}[t] {l}
\textbf{while} (x \sim c)~  \textbf{ do}  ~
 \{  x := \verb+f+(x); 
 \}  
\end{array}
\end{equation}
where $\sim \in \{<, \leq, >, \geq \}$, $c \in \mathbb{Z}$, 
and the statement $\verb+s+: x := \verb+f+(x);$ is a monotonic statement. Termination  of a program of the form (\ref{cyclesWithoutCond}) is very straightforward and can be decided using Lemma \ref{BoundeAboveBelow}.

\begin{lemma} \label{BoundeAboveBelow}

Let $\mathcal{L}$ be a program of the form (\ref{cyclesWithoutCond}). Then $\mathcal{L}$ is terminating if $\sim \in \{<, \leq \}$ and $\verb+s+ \uparrow$ or $\sim \in \{>, \geq \}$ and $\verb+s+ \downarrow$. On the other hand, $\mathcal{L}$ is non-terminating if $\sim \in \{<, \leq \}$ and $\verb+s+ \downarrow$ or  $\sim \in \{>, \geq \}$ and $\verb+s+ \uparrow$, given that the initial value of $x$ (let us call it $x_0$) satisfies the loop guard. 
\end{lemma}

\subsection{Single-path Loops with Diagonal Constraints} \label{sec: GapConst}

In this section we consider single-path loops with diagonal constraints of the form

\begin{equation} \label{cyclesWithGapConstraints}
 \begin{array}[t] {l}
\textbf{while} ~ ( (x - y) \sim c)~ \textbf{ do} ~ 
 \{  x := \verb+f+_1(x);   ~  y := \verb+f+_2(y); 
 \}  
\end{array}
\end{equation}
where $\sim \in \{<, \leq, >, \geq \}$, $c \in \mathbb{Z}$, and the statements $\verb+s+_1: x := \verb+f+_1(x);$ and $\verb+s+_2: y := \verb+f+_2(y);$ are monotonic statements.

In Table \ref{table:classification} we classify linear monotone expressions into regular and irregular expressions, where regular expressions are classified further into regular expressions with arithmetic progression and regular expressions with geometric progression.  We model the loop iteration update of each loop variable as a recurrence equation over  a new variable $n \in \mathbb{N}$ denoting the loop counter. 

\begin{table} 
 \centering
 \adjustbox{max width=\textwidth}{
\begin{tabular}{|c|c|c|c|}  
\hline 
Update expression & Regularity class &  Recurrence equation & Notation \\ 
\hline
$x:= x + v, v \neq 0 $ & Regular with arithmetic progression & $x_n = x_0 + v * n$ & $R_{a}$ \\ 
\hline 
$x:= u * x, u > 1 $ & Regular with geometric progression & $x_n = x_0 * u^{n}$ & $R_{g}$ \\ 
\hline
$x:= u* x + v, u > 1, v \neq 0 $ & Irregular &  $x_n = u^{n} * x_0 + \sum_{n = 0}^{n-1} (u ^{n} * v) $  & $I$ \\ 
\hline
\end{tabular} }
\caption{A summary of classification of linear monotone expressions} \label{table:classification}
\end{table}

Recurrence equations are fundamental models used for the definition of relations between consecutive elements of a sequence.  In this work, we show that for linear monotone programs, it is more natural to represent the behaviour of loop variables and to reason about termination using recurrence equations instead of using conventional approaches such as linear ranking approach.  Using such formalization, the analysis can take advantage of fundamental mathematical properties of various types of recurrence relations which arise during the analysis. As shown in Table \ref{table:classification} there are two types of recurrence equations that arise during the analysis of monotone programs: linear and exponential equations. 

We now discuss different procedures to analyse termination of a program of the form (\ref{cyclesWithGapConstraints}) depending on the various types of recurrence relations which arise during the analysis.
We consider the case where $\sim \in \{>, \geq \}$ since the analysis of the case where $\sim \in \{<, \leq \}$  is very similar.  Note that if the variables move in opposite directions (i.e. one increases and the other decreases) then the analysis is trivial (i.e. if $x$ moves up and $y$ moves down then the program is non-terminating, while if $x$ moves down and $y$ moves up then the program is terminating, given that the initial values $(x_0, y_0)$ satisfy the loop guard). However, if the variables move in the same direction then the analysis can be non-trivial.

\subsubsection{When both variables grow/decay linearly of the form $R_a$.} \label{sec: Ra}

Let $\verb+s+_1$ and $\verb+s+_2$ be monotonic statements of the form $R_{a}$. Termination of a program of the form (\ref{cyclesWithGapConstraints}) can be then decided using the following rules.

\begin{enumerate}



\item If $(\verb+s+_1 \uparrow \land ~ \verb+s+_2 \uparrow \land ~ (v_1 \geq v_2)) $ then the program is non-terminating. If $(\verb+s+_1 \uparrow \land ~ \verb+s+_2 \uparrow \land ~ (v_1 < v_2)) $ then the program is terminating.

\item If ($\verb+s+_1 \downarrow \land ~ \verb+s+_2 \downarrow \land ~ (|v_1| \leq |v_2|)) $ then the program is non-terminating. If $(\verb+s+_1 \downarrow \land ~ \verb+s+_2 \downarrow \land ~ (|v_1| > |v_2|)) $ then the program is terminating.

\end{enumerate}

\subsubsection{When both variables grow/decay exponentially of the form $R_{g}$.} \label{sec: Rg}

Let $\verb+s+_1$ and $\verb+s+_2$ be monotonic statements of the form  $R_{g}$. Termination of a program of the form (\ref{cyclesWithGapConstraints}) can be then decided using the following rules.

\begin{enumerate}

\item If $(\verb+s+_1 \uparrow \land ~ \verb+s+_2 \uparrow \land ~ (u_1 \geq u_2)) $ then the program is non-terminating. If $(\verb+s+_1 \uparrow \land ~\verb+s+_2 \uparrow \land ~ (u_1 < u_2)) $ then the program is terminating.

\item If ($\verb+s+_1 \downarrow \land ~ \verb+s+_2 \downarrow \land ~ (|u_1| \leq |u_2|)) $ then the program is non-terminating. If  $(\verb+s+_1 \downarrow \land ~ \verb+s+_2 \downarrow \land ~ (|u_1| > |u_2|)) $ then the program is terminating.

\end{enumerate}

\subsubsection{When one variable grows/decays linearly and the other exponentially.} \label{sec: RaANDRg}

Let $\verb+s+_1$  be a monotonic statement of the form $R_{a}$ and  $\verb+s+_2$  be a monotonic statement of the form $R_g$ or $I$. Then termination can be decided using the following rules. 

\begin{enumerate}

\item If $(\verb+s+_1 \uparrow \land ~ \verb+s+_2 \uparrow)$ then the program is terminating since for any exponential function and any linear function with positive growth, the exponential function will eventually outstrip the linear function.

\item If ($\verb+s+_1 \downarrow \land ~ \verb+s+_2 \downarrow ) $. This case requires a special decision procedure (see Algorithm \ref{alg:specialProcedureCase4})  since it is not obvious whether there will be an iteration at which the guard  $(x_n - y_n \sim c)$ can be violated.
\end{enumerate}

\begin{algorithm} 
\caption{Special decision procedure for case 2}
\label{alg:specialProcedureCase4}
\textbf{Input}:  $(x_0, y_0, v_1, v_2, u, c, \sim)$ \\
\textbf{output}: \{\textit{terminating, non-terminating}\}\\
\textbf{int} $n := 1$; \\
\textbf{while} (\textit{true}) \\ 
\hspace*{10 pt}\{ \\
\hspace*{15 pt} $x_n = (x_0 + v_1 * n)$; \hspace*{5 pt} $y_n = u^{n} * y_0 + \sum_{n = 0}^{n-1} (u ^{n} * v_2)$;  \\
\hspace*{15 pt} \textbf{if} ( ($\sim= ` > $' $\land$ $y_n \geq (x_n - c)$) $\lor$ ($\sim= ` \geq $' $\land$ $y_n > (x_n - c)$) ) \textbf{return} \textit{terminating}; \\
\hspace*{15 pt} \textbf{if} $(y_n < (x_n - c) \land x_n < 0 \land y_n <0)$ \textbf{return} \textit{non-terminating};\\
\hspace*{15 pt}  $n:= n+1$; \\
\hspace*{10 pt} \}
\end{algorithm} 


Note that when  algorithm \ref{alg:specialProcedureCase4} terminates with `terminating' then the search has reached an iteration where the loop guard is violated. On the other hand, when the algorithm terminates with `non-terminating' then the search has reached an iteration where $(y_n < (x_n - c) \land x_n < 0 \land y_n <0)$. This condition is sufficient to guarantee non-termination of this form of loops since $x$ is decreasing linearly while $y$ is decreasing exponentially.

We now turn to discuss the case where $x$ grows/decays exponentially and $y$ grows/decays linearly. In this case termination can be decided as follows.

\begin{enumerate}

\item If $(\verb+s+_1 \downarrow \land ~ \verb+s+_2\downarrow)$ then the program is terminating since there will be an iteration where $x$ can be smaller than $y$. This is due to the assumption that $x$ is decreasing exponentially while  $y$ is decreasing linearly.

\item If ($\verb+s+_1 \uparrow \land ~ \verb+s+_2 \uparrow ) $. This case requires a special decision procedure (similar to Algorithm \ref{alg:specialProcedureCase4} above) but it differs at the condition at which it returns `non-terminating' which will be of the form $(x_n > (y_n + c)) $. This is due to the observation that  once the exponential function outstrips the linear function at some iteration it continues to outstrip it at all future iterations.
\end{enumerate}

\subsubsection{When both variables grow/decay exponentially.} \label{sec: RgANDI}

In this case a decision procedure similar to Algorithm \ref{alg:specialProcedureCase4} can be used to decide termination. The procedure differs from Algorithm \ref{alg:specialProcedureCase4} at the condition at which it terminates with `non-terminating'. That is, if both variables are moving up then the procedure terminates with `non-terminating' when the search reaches an iteration at which ($(x_n > y_n +  c) \land u_1 \geq u_2$) since both $x$ and $y$ grow exponentially as the loop proceeds. On the other hand, if both variables are moving down then the condition at which the algorithm terminates with `non-terminating' will be of the form $(y_n < (x_n -c) \land x_n <0 \land y_n < 0 \land u_1 \leq u_2)$ since both variables decay exponentially as the loop proceeds.

\paragraph*{}

In Table \ref{table:classificationStoppingCond} we summarise the stopping condition at which a decision procedure of the form \ref{alg:specialProcedureCase4} returns `non-terminating' when the variables $x$ and $y$ move in the same direction but with different mathematical behaviour. Note that the procedures return `terminating' when the loop guard $(x - y \sim c)$, where $\sim \in \{>, \geq\}$, is violated but they differ at the stopping condition at which they return `non-terminating' since this depends on the characteristics of the recurrence update equations of $x$ and $y$ (i.e. whether they grow/decay linearly or exponentially). The +, -- signs used as a superscript of notations $R_a$, $R_g$, and $I$ to indicate whether the variable is moving up (+) or moving down (--).

\begin{table} 
 \centering
 \small
\begin{tabular}{|c|c|}  
\hline 
Update statements & Stopping condition (non-termination) \\ 
\hline
$(1) ~ \verb+s+_1 = R_a^{-} \land \verb+s+_2 = R_g^{-}$ & \multirow{ 2}{*} {$(y_n < (x_n - c) \land x_n < 0 \land y_n <0)$} \\ (2) $\verb+s+_1 = R_a^{-} \land \verb+s+_2 = I^{-}$& \\
\hline
$(3) ~ \verb+s+_1 = R_g^{+} \land \verb+s+_2 = R_a^{+}$ & \multirow{ 2}{*} {$(x_n > (y_n + c)) $} \\ (4) $\verb+s+_1 = I^{+} \land \verb+s+_2 = R_a^{+}$ & \\
\hline
$(5) ~\verb+s+_1 = R_g^{+} \land \verb+s+_2 = I^{+}$ & \multirow{ 2}{*} {$( (x_n > y_n +  c) \land u_1 \geq u_2)$} \\ (6) $\verb+s+_1 = I^{+} \land \verb+s+_2 = R_g^{+}$ & \\
\hline
$(7)~ \verb+s+_1 = R_g^{-} \land \verb+s+_2 = I^{-}$ & \multirow{ 2}{*} {$(y_n < (x_n -c) \land x_n <0 \land y_n < 0 \land u_1 \leq u_2)$} \\ (8) $\verb+s+_1 = I^{-} \land \verb+s+_2 = R_g^{-}$ & \\
\hline
\end{tabular} 
\caption{Stopping condition at which the procedure returns non-terminating} \label{table:classificationStoppingCond}
\end{table}

\vspace*{-2em}




\section{Multi-path Loops with Diagonal-free Constraints}

Paths of multi-path loops can interleave in a non-trivial manner. Reasoning about termination of  multi-path loops is therefore a challenging task. The class of multi-path loops that we consider has the following form

\begin{equation} \label{cyclesWithCond1}
\textbf{while} ~ (x \sim c) ~  \textbf{ do}  ~ \{  \textbf{if} ~ (x \sim^{'} c_1) ~ \textbf{then} ~ x := \verb+f+_1(x); ~ \textbf{else } ~ x := \verb+f+_2(x); \}
 \end{equation}
 
\paragraph*{•} where the statements $\verb+s+_1:x := \verb+f+_1(x);$ and $\verb+s+_2:x := \verb+f+_2(x);$ are monotonic statements, $\sim, \sim^{'} \in \{<, \leq, >, \geq \}$, and $c, c_1 \in \mathbb{Z}$. We refer to the assignment $x := \verb+f+_1(x)$ as a conditional assignment, as its execution depends on the truth value of  the condition $(x \sim^{'} c_1)$. As we shall see in this section the complexity of termination analysis of a program of the form (\ref{cyclesWithCond1}) varies depending on the class of monotonicity of the statements  $\verb+s+_1$ and $\verb+s+_2$ and whether the conditional branches of the program interleave between iterations. During the analysis of the different cases of a program of the form (\ref{cyclesWithCond1}) we use $\phi$ to refer to the loop guard $(x \sim c)$ and $\verb+B+$ to refer to the condition $(x \sim^{'} c_1)$. We write $\sim_{\phi}$ to refer to the relation operator used in the loop guard, $\sim_{\verb+B+}$ to refer to the relation operator used in the if-branch, and $\sim_{\neg \verb+B+}$ to refer to the relation operator used in the else-branch.

\begin{observation} \label{Obs1}
If the statements $\verb+s+_1$ and $\verb+s+_2$ in program (\ref{cyclesWithCond1}) have the same monotonicity behaviour (i.e. both are monotonically increasing or monotonically decreasing) then we can safely ignore \verb+B+ and use Lemma \ref{BoundeAboveBelow} to verify termination.
\end{observation}

\subsection{When One Statement is Monotone and the other is Constant} \label{sec:f1Monotonef2Constant}

In this section we analyse termination of a program of the form (\ref{cyclesWithCond1}) under the assumption that one of the update statements is strictly monotone and the other is constant. In this case there are several sub-cases to consider depending on the operators $\sim$, $\sim^{'}$ and the class of monotonicity of the statements $\verb+s+_1$ and $\verb+s+_2$.

\begin{enumerate}

\item When $\sim_{\phi} \in \{ >, \geq\}$ (i.e. $x$ is bounded from below in loop guard $\phi$) and $\sim_{\verb+B+} \in \{<, \leq\}$ (i.e. $x$ is bounded from above in $\verb+B+$) and $\verb+s+_1$  $\uparrow$  and $\verb+s+_2$  $\rightarrow_{b}$ (i.e. $\verb+f+_2(x) : = b$, where $b$ is a constant). From the syntactic structure of the given loop form and by observing the way the loop variable $x$ is changed w.r.t to the loop condition it is easy to see that for this form of loops there is only one condition of non-termination. This happens when $(b \sim_{\phi} c)$, where $\sim_{\phi}$ is the relational operator used in the guard $\phi$. Note that the if-branch can not contribute towards the termination of the loop since $\verb+s+_1$ is monotonically increasing and $x$ is bounded from below in $\phi$. However, from the given loop form we note that when the if-branch is triggered then the else-branch will be triggered at some later iteration.  Hence, the formula of non-termination for this form of loops can be described as follows
\begin{equation} \label{caseOne}
NT = (x_0 \models \phi \land  (b \sim_{\phi} c) )
\end{equation}

where $x_0$ denotes the initial value of $x$ and $x_0 \models \phi$ means that the guard $\phi$ is evaluated to \textit{true} when $x = x_0$. If formula NT is satisfiable then we say the loop is non-terminating. Note that formula (\ref{caseOne}) can be used also to verify termination of loops in which $\sim_{\phi} \in \{<, \leq\}$, $\sim_{\verb+B+} \in \{ >, \geq\}$, $\verb+s+_1$  $\downarrow$, and $\verb+s+_2$  $\rightarrow_{b}$ for the same reasoning given above.

\item When $\sim_{\phi} \in \{ >, \geq\}$ and $\sim_{\verb+B+}  \in \{ <, \leq\}$  and $\verb+s+_1 \downarrow$ and $\verb+s+_2 \rightarrow_{b}$. For this form of loops there is only one case of non-termination. This case happens when the else-branch is triggered during the execution of the loop and that $(b \sim_{\phi} c)$ and  $(b \sim_{\neg \verb+B+}  c_1)$. Note that if the if-branch is triggered during the execution of the loop then the loop will eventually terminate since $x$ is bounded from below in $\phi$ and from above in $\verb+B+$ and $\verb+s+_1$ is monotonically decreasing.
\begin{equation} \label{caseTwo}
NT = (x_0 \models \phi \land x_0 \not\models  \verb+B+ \land  (b \sim_{\phi} c)  \land  (b \sim_{\neg \verb+B+} c_1) )
\end{equation}

Formula (\ref{caseTwo}) can be used also to verify termination of loops in which $\sim_{\phi} \in \{<, \leq\}$,  $\sim_{\verb+B+}  \in \{>, \geq \}$,  $\verb+s+_1 \uparrow$, and  $\verb+s+_2 \rightarrow_{b}$ for the same reasoning given above.

\item When  $\sim_{\phi} \in \{ <, \leq\}$ and $\sim_{\verb+B+}  \in \{ <, \leq\}$ and $\verb+s+_1 \uparrow$ and $\verb+s+_2 \rightarrow_{b}$.
Analysing termination of this form of loops is non-trivial since the conditional branches of the loop can interleave between iterations. However, before discussing the conditions of non-termination of this form of loops we need to observe the following. From the syntactic structure of the given loop form we note that when the loop is non-terminating and the if-branch is triggered then the else-branch will be triggered at some later iteration since $x$ is bounded from above in \verb+B+  and $\verb+s+_1$ is monotonically increasing. On the other hand, when the else-branch is triggered then the if-branch may or may not be triggered at future iterations depending on the value of the constant $b$.  Hence, finding the conditions of non-termination of this form of loops requires some numerical analysis.	 
Since $x$ is bounded from above in $\verb+B+$ and $\verb+s+_1$ is monotonically increasing then we need to compute the smallest value of $x$ that falsifies the condition \verb+B+ when the if-branch is triggered with $x = d$, where $d \in \{x_0, b\}$. That value can be computed using some special numerical procedures depending on the class of monotonicity of the statement $\verb+s+_1$.  That is, if $\verb+s+_1$ is $R_{a}$ (i.e. has the form $x:= x + v$)  then one can use formula (\ref{maxequation}). On the other hand, if $\verb+s+_1$ is $R_{g}$ or $I$ then one needs to use Algorithm \ref{alg:specialProcedureCasePsiGAndI}, where the bound $c_1$ that appears in the algorithm is the bound that $x$ is compared to in the condition \verb+B+. 

\begin{equation} \label{maxequation}
\psi_{a}(d) = 
\begin{cases}
 ((c_1 + v) - ( (c_1 - d) ~ \% ~ v)) & \textrm{if $ \sim_{\textsc{B}} =`\leq$' } \\
( ( (c_1  - 1) +  v) - ( (c_1 - 1) -  d) ~ \% ~  v) )) & \textrm{if $ \sim_{\textsc{B}} =`<$' }
\end{cases}
\end{equation} 

Let us explain the intuition behind formula (\ref{maxequation}). Note that at each iteration the if-branch is triggered the variable $x$ is  incremented by $v$. Hence, the maximum value that $x$ can reach from the consecutive execution of 
the if-branch is either $(c_1+ v)$ or $((c_1  - 1) +  v)$ depending on $\sim_{\verb+B+}$. Therefore, if $(c_1 - d)$ is a multiple of $v$ then the smallest value of $x$ that falsifies \verb+B+ will be either $(c_1+ v)$ or $(c_1  - 1) +  v)$. On the other hand, if $(c_1 - d)$ is not a multiple of $v$ then the remainder needs to be subtracted from the maximum value $(c_1+ v)$ or $((c_1  - 1) +  v)$ depending on $\sim_{\verb+B+}$.

\begin{algorithm} 
\caption{Special procedure $\psi_{g}, \psi_{I}$}
\label{alg:specialProcedureCasePsiGAndI}
\textbf{Input}:  $(d, c_1, u, v, \sim_{\verb+B+})$ \\
\textbf{int} $n := 1$; $x_{n-1} := d$; \\
\textbf{while} (\textit{true}) \\ 
\hspace*{10 pt}\{ \\
\hspace*{20 pt} $x_n =  (u * x_{n-1} + v)$;  \hspace*{5 pt} $n++$;\\
\hspace*{20 pt} \textbf{if} ( ($\sim_{\verb+B+}= ` < $' $\land$ $x_n \geq c_1$) $\lor$ ($\sim_{\verb+B+}= ` \leq $' $\land$ $x_n > c_1$) ) \textbf{return} $x_n$; \\
\hspace*{10 pt} \}
\end{algorithm}

By analysing the conditional branches of the loop and their possible interleaving one can see that there are four conditions of non-terminating for this form of loops, which can be formalised as follows

\begin{equation} \label{caseFive}
NT = \begin{array}[t]{l} 
   ( (x_0 \models \phi \land x_0 \not \models \verb+B+ \land b \sim_{\phi} c \land b \sim_{\neg \verb+B+} c_1)   ~\lor \\  ~ (x_0 \models \phi \land x_0 \models \verb+B+ \land \psi(x_0) \sim_{\phi} c \land b \sim_{\phi} c \land b \sim_{\neg \verb+B+} c_1) ~\lor \\
   ~(x_0 \models \phi \land x_0 \models \verb+B+ \land \psi(x_0) \sim_{\phi} c \land b \sim_{\phi} c \land b \sim_{\verb+B+} c_1 \land \psi(b) \sim_{\phi} c) ~\lor \\ 
~  (x_0 \models \phi \land x_0 \not \models \verb+B+ \land b \sim_{\phi} c \land b \sim_{\verb+B+} c_1 \land \psi(b) \sim_{\phi} c)  )
\end{array}
\end{equation}
where $\psi(x_0)$ returns the smallest value of $x$ at which the if-branch becomes disabled after being triggered using the value $x_0$, and $\psi(b)$  returns the smallest value of $x$ at which the if-branch becomes disabled after being triggered using the value $b$.  Note that we write $\psi(x)$ in the formula which can be either $\psi_a$ or $\psi_g$ or $\psi_I$ depending on the class of monotonicity of the statement $\verb+s+_1$.

\item When $\sim_{\phi} \in \{ <, \leq\}$ and $\sim_{\verb+B+}  \in \{ <, \leq\}$ and $\verb+s+_1 \downarrow$ and $\verb+s+_2 \rightarrow_{b}$.  For this form of loops there are two cases of non-termination. The first case is when  the if-branch is triggered during the execution of the loop. In this case the loop will not terminate since $x$  is bounded from above in both $\phi$ and $\verb+B+$ and $\verb+s+_1$ is monotonically decreasing.  The second case is when the else-branch is triggered and that $(b \sim_{\phi} c)$. In this case the loop will not terminate regardless of which branch is triggered between iterations. 
\begin{equation} \label{caseSix}
NT = ( (x_0 \models \phi \land x_0  \models \verb+B+) \lor (x_0 \models \phi \land x_0  \not \models \verb+B+ \land b \sim_{\phi} c) )
\end{equation}

Formula (\ref{caseSix}) can be used also to verify termination of loops in which $\sim_{\phi} \in \{ >, \geq \}$ and $\sim_{\verb+B+} \in \{ >, \geq \}$ and $\verb+s+_1 \uparrow$ and $\verb+s+_2 \rightarrow_{b}$.

\item When $\sim_{\phi} \in \{ >, \geq\}$ and $\sim_{\verb+B+}  \in \{  >, \geq \}$ and $\verb+s+_1 \downarrow$ and $\verb+s+_2 \rightarrow_{b}$. Termination analysis of this form of loops is similar to case 3 above in the sense that deriving the conditions of non-termination requires performing some numerical analysis. The only difference here is the formula that is used to compute the smallest value of $x$ that falsifies the condition \verb+B+ when the if-branch is triggered at some prior iteration with $x =d$, where $d \in \{x_0, b\}$, since $\verb+s+_1$ in this case is monotonically decreasing. 
This value can be computed using formula (\ref{minequation}) in case $\verb+s+_1$ is $R_{a}$ (i.e. has the form $x:= x + v$) or using a special procedure similar to Algorithm \ref{alg:specialProcedureCasePsiGAndI} in case $\verb+s+_1$ is $R_{g}$ or $I$.
\begin{equation} \label{minequation}
\psi^{'}_{a} (d) = 
\begin{cases}
( (c_1 -  v) + ( (d - c_1) ~ \% ~ v)) & \textrm{if $ \sim_{\textsc{B}} = `\geq$'} \\
 ( ((c_1 +1) -  v) + ( (d - (c_1 + 1)) ~ \% ~ v))) & \textrm{if $  \sim_{\textsc{B}} =`>$'}
\end{cases}
\end{equation}

Similar to case 3 above there are four conditions of non-termination of this form of loops which can be formalized as given in formula (\ref{caseSeven}).
\begin{equation} \label{caseSeven}
NT = \begin{array}[t]{l} 
   ( (x_0 \models \phi \land x_0 \not \models \verb+B+ \land b \sim_{\phi} c \land b \sim_{\neg \verb+B+} c_1)   ~\lor \\  ~ (x_0 \models \phi \land x_0 \models \verb+B+ \land \psi^{'}(x_0) \sim c \land b \sim_{\phi} c \land b \sim_{\neg \verb+B+} c_1 ) ~\lor \\
   ~ (x_0 \models \phi \land x_0 \models \verb+B+ \land \psi^{'}(x_0) \sim_{\phi} c \land b \sim c \land b \sim_{\phi} c_1 \land \psi^{'}(b) \sim_{\phi} c) ~ \lor \\ 
~  (x_0 \models \phi \land x_0 \not \models \verb+B+ \land b \sim_{\phi} c \land b \sim_{\verb+B+} c_1 \land \psi^{'}(b) \sim_{\phi} c)  )
\end{array}
\end{equation}

\end{enumerate}

\subsection{When Both Update Statements are Strictly Monotone} \label{sec:ConditioalBothMonotone}

This is the most challenging form of loops to consider since at some iteration of the loop the variable $x$ may be updated using the arbitrary monotone statement $\verb+s+_1$ while in some later iteration it may be updated using the arbitrary monotone statement $\verb+s+_2$. Hence, to analyse termination of this form of loops  we need to consider several sub-cases depending on the class of monotonicity of the statements $\verb+s+_1$ and  $\verb+s+_2$ and the operators $\sim$ and $\sim^{'}$.  
\begin{enumerate}

\item When $\sim_{\phi} \in \{ >, \geq\}$ and $\sim_{\verb+B+}  \in \{  >, \geq \}$ and $\verb+s+_1 \uparrow$ and $\verb+s+_2 \downarrow$. Surprisingly, this case is trivial since there is only one condition under which the loop can be non-terminating as described in formula (\ref{caseOneMonotone}). 
\begin{equation} \label{caseOneMonotone}
NT = (x_0 \models \phi \land x_0 \models \verb+B+)
\end{equation}

That is, if the if-branch is triggered in the first iteration of the loop then the loop is non-terminating since $\verb+s+_1$ is monotonically increasing and $x$ is bounded from below in both $\phi$ and $\verb+B+$. Note that the else-branch cannot contribute towards the non-termination of the loop since $\verb+s+_2$ is monotonically decreasing and $x$ is bounded from below in $\phi$ and from above in $\neg \verb+B+$.
Formula (\ref{caseOneMonotone}) can be used also to verify termination of loops in which $\sim_{\phi} \in \{ <, \leq\}$, $\sim_{\verb+B+}  \in \{  <, \leq \}$, $\verb+s+_1 \downarrow$, and $\verb+s+_2 \uparrow$ for the same reasoning given above.

\item When $\sim_{\phi} \in \{ <, \leq\}$ and $\sim_{\verb+B+}  \in \{  <, \leq \}$ and $\verb+s+_1 \uparrow$ and $\verb+s+_2 \downarrow$. For this form of loops it is not possible to derive a concrete formula for verifying termination as we have done for the previous cases since in case of non-termination the two branches will activate each other in alternation and hence the loop variable will be updated using any of the two arbitrary statements.  We therefore use a special procedure to analyse termination of this form of loops. The procedure (see Algorithm \ref{alg:specialProcedure4.3.2}) is guaranteed to return with a correct answer. The procedure uses the auxiliary formulas $\psi(x)$ and $\psi^{'}(x)$ to compute the precise value of $x$ between iterations. 
The procedure maintains a list called PASSED to store the values of $x$ that are already examined. Note that if at some iteration of the loop the guard $\phi$ is violated then the loop is terminating. On the other hand, if the search reaches a fixed point (i.e. the same value of $x$ is encountered again during the search) then the loop is non-terminating. Reaching a fixed point in case of non-termination is guaranteed since the domain of $x$ at both branches is finite. Note also that for this form of loops the if-branch is the only branch that can contribute towards the termination of the loop as $x$ is bounded from above in  $\phi$ and $\verb+s+_1$ is monotonically increasing. This is the reason why we  check  for termination only at the if-branch. 

\begin{algorithm} [h!]
\caption{Special decision procedure for case 2 at Section \ref{sec:ConditioalBothMonotone}}
\label{alg:specialProcedure4.3.2}
\textbf{Input}:  $(x_0, \phi, \verb+B+)$ \\
\textbf{output}: \{\textit{terminating, non-terminating}\}\\
PASSED := $\emptyset$;  \textbf{int} VAL = $x_0$; \\
\textbf{while} (\textit{true}) \\
\hspace*{10 pt} \textbf{if} (VAL $\not \models \verb+B+$) \textbf{then} \textbf{if} ($\psi^{'}$(VAL) = $x_i$) for any $x_i \in $ PASSED    \textbf{then} \\ \hspace*{20 pt} \textbf{return} \textit{`non-terminating'} \textbf{else} \{add VAL to PASSED; VAL:= $\psi{'}$(VAL)\} \\ 
\hspace*{10 pt} \textbf{if} (VAL $\models \verb+B+$) \textbf{then} \textbf{if} ($\psi$(VAL) = $x_i$) for any $x_i \in $ PASSED    \textbf{then} \\ \hspace*{20 pt} \textbf{return} \textit{`non-terminating'} \textbf{else if} ($\psi$(VAL) $\not \models \phi$) \textbf{then} \\ \hspace*{20 pt}    \textbf{return} \textit{`terminating'} \textbf{else} \{add VAL to PASSED; VAL:= $\psi$(VAL)\} 
\end{algorithm}

\item  When $\sim_{\phi} \in \{ <, \leq\}$ and $\sim_{\verb+B+}  \in \{>, \geq\}$ and $\verb+s+_1 \uparrow$ and $\verb+s+_2 \downarrow$. For this form of loops there is only one case of non-termination. This case happens when the else-branch is triggered at the first iteration of the loop. Since $x$ is bounded from above in $\phi$ and from above in $\neg \verb+B+$ and $\verb+s+_2$ is  monotonically decreasing then the loop will not terminate in this case. Note that the if-branch cannot contribute towards the non-termination of the loop since $x$ is bounded from below in \verb+B+ and from above in $\phi$ and $\verb+s+_1$ is  monotonically increasing.

\begin{equation} \label{caseThreeMonotone}
NT = (x_0 \models \phi \land x_0 \not\models \verb+B+)
\end{equation}

Note that formula (\ref{caseThreeMonotone}) can be used also to verify termination of loops in which $\sim_{\phi} \in \{>, \geq\}$ and $\sim_{\verb+B+}  \in \{ <, \leq\}$ and $\verb+s+_1 \downarrow$ and $\verb+s+_2 \uparrow$.

\item  When $\sim_{\phi} \in \{>, \geq\}$  and $\sim_{\verb+B+}  \in \{ <, \leq\}$ and $\verb+s+_1 \uparrow$ and $\verb+s+_2 \downarrow$. This case is similar to case 2 above since in case of non-termination the conditional branches will interleave between iterations in a non-trivial manner. The only difference here is that the if-branch in this case cannot contribute towards the termination of the loop while the else-branch can do so. 

\begin{algorithm} [h!]
\caption{Special decision procedure for case 4 at Section \ref{sec:ConditioalBothMonotone}}
\label{alg:specialProcedure4.3.4}
\textbf{Input}:  $(x_0, \phi, \verb+B+)$ \\
\textbf{output}: \{\textit{terminating, non-terminating}\}\\
PASSED := $\emptyset$;  \textbf{int} VAL = $x_0$; \\
\textbf{while} (\textit{true}) \\
\hspace*{10 pt} \textbf{if} (VAL $  \models \verb+B+$) \textbf{then} \textbf{if} ($\psi$(VAL) = $x_i$) for any $x_i \in $ PASSED    \textbf{then} \\ \hspace*{20 pt} \textbf{return} \textit{`non-terminating'} \textbf{else} \{add VAL to PASSED; VAL := $\psi$(VAL) \} \\  
\hspace*{10 pt} \textbf{if} (VAL $ \not\models \verb+B+$) \textbf{then} \textbf{if} ($\psi{'}$(VAL) = $x_i$) for any $x_i \in $ PASSED    \textbf{then} \\ \hspace*{20 pt} \textbf{return} \textit{`non-terminating'} \textbf{else if} ($\psi{'}$(VAL) $\not \models \phi$) \textbf{then} \\ \hspace*{20 pt}    \textbf{return} \textit{`terminating'} \textbf{else} \{add  VAL to PASSED; VAL := $\psi{'}$(VAL);\} 
\end{algorithm} 
 \vspace*{-1em}

\end{enumerate}

In Table \ref{table:verification} we summarize the termination rules of all possible cases of multi-path loops with diagonal-free constraints of the form (\ref{cyclesWithCond1}), where $c$ represents the bound that the loop variable is compared to in $\phi$, and $c_1$ is the bound that the variable is compared to in the condition \verb+B+. Since the relational operator used in the guard $\phi$ and in the condition $\verb+B+$ can be either bounded from below $\{>, \geq\}$ or bounded from above $\{<, \leq\}$ and that the update statements $\verb+s+_1$ and $\verb+s+_2$ can be either $\uparrow$ or $\downarrow$ or $\rightarrow_{b}$, then we have a total of 36 cases to consider as shown in the table. We now discus some simple examples of loop programs to demonstrate how the termination rules described in Table \ref{table:verification} can be used to decide termination. 

\begin{table} [h!]
 \centering
\adjustbox{max width=\textwidth}{
\begin{tabular}{|c|c|} 
\hline 
Loop form & Termination rule \\ 
\hline
 (1) $(\sim_{\phi} \in \{>, \geq\} \land \sim_{\verb+B+}  \in \{<, \leq\} \land \verb+s+_1 \uparrow \land \verb+s+_2 \rightarrow_{b})  $ & \multirow{2}{*}{$(x_0 \models \phi \land  (b \sim_{\phi} c) ) $}  \\ (2) $(\sim_{\phi} \in \{<, \leq\} \land \sim_{\verb+B+}  \in \{ >, \geq\} \land \verb+s+_1  \downarrow \land \verb+s+_2 \rightarrow_{b})$  &\\ (3) $(\sim_{\phi} \in \{>, \geq\} \land \sim_{\verb+B+}  \in \{ >, \geq\} \land \verb+s+_1  \rightarrow_{b} \land \verb+s+_2 \uparrow)$ &\\ (4) $(\sim_{\phi} \in \{<, \leq\} \land \sim_{\verb+B+}  \in \{ <, \leq\} \land \verb+s+_1  \rightarrow_{b} \land \verb+s+_2 \downarrow)$ & \\
\hline

(5) $(\sim_{\phi} \in \{ >, \geq\} \land \sim_{\verb+B+}  \in \{ <, \leq\}  \land \verb+s+_1 \downarrow \land \verb+s+_2 \rightarrow_{b})$ & \multirow{ 2}{*}{$ (x_0 \models \phi \land x_0 \not\models  \textsc{B} \land  (b \sim_{\phi} c)  \land  (b \sim_{\neg \textsc{B}}  c_1) ) $} \\ (6) $(\sim_{\phi} \in \{<, \leq\} \land \sim_{\verb+B+}  \in \{>, \geq \} \land \verb+s+_1 \uparrow \land \verb+s+_2 \rightarrow_{b})$ & \\ 
\hline

(7) $(\sim_{\phi} \in \{ >, \geq\} \land \sim_{\verb+B+}  \in \{ >, \geq\}  \land \verb+s+_1 \rightarrow_{b} \land \verb+s+_2 \downarrow )$ & \multirow{ 2}{*}{$ (x_0 \models \phi \land x_0 \models  \textsc{B} \land  (b \sim_{\phi} c)  \land  (b \sim_{\textsc{B}}  c_1) ) $} \\ (8) $(\sim_{\phi} \in \{<, \leq\} \land \sim_{\verb+B+}  \in \{<, \leq \} \land  \verb+s+_1 \rightarrow_{b} \land \verb+s+_2 \uparrow )$ & \\ 
\hline

(9) $(\sim_{\phi} \in \{ >, \geq\} \land \sim_{\verb+B+}  \in \{ <, \leq\}  \land \verb+s+_1 \rightarrow_{b} \land \verb+s+_2 \uparrow)$ & \multirow{ 2}{*}{$  \begin{array}[t]{l}  ( (x_0 \models \phi \land x_0  \not\models \textsc{B}) ~ \lor  ~(x_0 \models \phi \land x_0  \models \textsc{B} \land b \sim_{\phi} c) ) \end{array} $}  \\ (10) $(\sim_{\phi} \in \{ <, \leq\} \land \sim_{\verb+B+}  \in \{ >, \geq\}  \land \verb+s+_1 \rightarrow_{b}	 \land \verb+s+_2 \downarrow)$ & \\

\hline

(11) $(\sim_{\phi} \in \{ <, \leq\} \land \sim_{\verb+B+}  \in \{ <, \leq\} \land \verb+s+_1 \downarrow \land \verb+s+_2   \rightarrow_{b})$ 
 & \multirow{ 2}{*}{$ \begin{array}[t]{l}  ( (x_0 \models \phi \land x_0  \models \textsc{B}) ~ \lor  ~(x_0 \models \phi \land x_0  \not \models \textsc{B} \land b \sim_{\phi} c) ) \end{array}$}  \\ (12) $(\sim_{\phi} \in \{ >, \geq \} \land \sim_{\verb+B+}  \in \{ >, \geq \} \land \verb+s+_1 \uparrow \land \verb+s+_2 \rightarrow_{b})$ & \\ 
\hline

 (13) $(\sim_{\phi} \in \{ <, \leq\} \land \sim_{\verb+B+}  \in \{ <, \leq\} \land \verb+s+_1   \uparrow \land \verb+s+_2  \rightarrow_{b})$ &  $\begin{array}[t]{l} 
   ( (x_0 \models \phi \land x_0 \not \models \textsc{B} \land b \sim_{\phi} c \land b \sim_{\neg \textsc{B}}  c_1)   ~\lor \\  ~ (x_0 \models \phi \land x_0 \models \textsc{B} \land \psi(x_0) \sim_{\phi} c \land b \sim_{\phi} c \land b \sim_{\neg\textsc{B}}  c_1) ~\lor \\
   ~(x_0 \models \phi \land x_0 \models \textsc{B} \land \psi(x_0) \sim c \land b \sim_{\phi} c \land b \sim_{\textsc{B}}  c_1 \land \psi(b) \sim_{\phi} c) ~\lor \\ 
~  (x_0 \models \phi \land x_0 \not \models \textsc{B} \land b \sim_{\phi} c \land b \sim_{\textsc{B}}  c_1 \land \psi(b) \sim_{\phi} c)  )
\end{array}$    \\ 
\hline

(14) $ (\sim_{\phi} \in \{ <, \leq\} \land \sim_{\verb+B+}  \in \{ >, \geq\}  \land \verb+s+_1 \rightarrow_{b} \land \verb+s+_2 \downarrow)$ &   $\begin{array}[t]{l} 
   ( (x_0 \models \phi \land x_0  \models \textsc{B} \land b \sim_{\phi} c \land b \sim_{ \textsc{B}}  c_1)   ~\lor \\  ~ (x_0 \models \phi \land x_0 \not \models \textsc{B} \land \psi(x_0) \sim_{\phi} c \land b \sim_{\phi} c \land b \sim_{\textsc{B}}  c_1) ~\lor \\
   ~(x_0 \models \phi \land x_0 \not \models \textsc{B} \land \psi(x_0) \sim c \land b \sim_{\phi} c \land b \sim_{\neg \textsc{B}}  c_1 \land \psi(b) \sim_{\phi} c) ~\lor \\ 
~  (x_0 \models \phi \land x_0 \models \textsc{B} \land b \sim_{\phi} c \land b \sim_{\neg \textsc{B}}  c_1 \land \psi(b) \sim_{\phi} c)  )
\end{array}$    \\ 
\hline

(15) $ (\sim_{\phi} \in \{ >, \geq\} \land \sim_{\verb+B+}  \in \{  >, \geq \} \land \verb+s+_1  \downarrow \land \verb+s+_2  \rightarrow_{b})$ & $\begin{array}[t]{l} 
   ( (x_0 \models \phi \land x_0 \not \models \textsc{B} \land b \sim_{\phi} c \land b \sim_{\neg \textsc{B}}  c_1)   ~\lor \\  ~ (x_0 \models \phi \land x_0 \models \textsc{B} \land \psi^{'}(x_0) \sim_{\phi} c \land b \sim_{\phi} c \land b \sim_{\neg \textsc{B}}  c_1 ) ~\lor \\
   ~ (x_0 \models \phi \land x_0 \models \textsc{B} \land \psi^{'}(x_0) \sim_{\phi} c \land b \sim_{\phi} c \land b \sim_{\phi} c_1 \land \psi^{'}(b) \sim_{\phi} c) \lor \\ 
~  (x_0 \models \phi \land x_0 \not \models \textsc{B} \land b \sim_{\phi} c \land b \sim_{\textsc{B}} c_1 \land \psi^{'}(b) \sim_{\phi} c)  )
\end{array}$ \\ 
\hline

(16) $ (\sim_{\phi} \in \{ >, \geq\} \land \sim_{\verb+B+}  \in \{<, \leq\}  \land \verb+s+_1 \rightarrow_{b} \land \verb+s+_2 \downarrow)$ &  $\begin{array}[t]{l} 
   ( (x_0 \models \phi \land x_0  \models \textsc{B} \land b \sim_{\phi} c \land b \sim_{ \textsc{B}}  c_1)   ~\lor \\  ~ (x_0 \models \phi \land x_0 \not \models \textsc{B} \land \psi^{'}(x_0) \sim_{\phi} c \land b \sim_{\phi} c \land b \sim_{\textsc{B}}  c_1) ~\lor \\
   ~(x_0 \models \phi \land x_0 \not \models \textsc{B} \land \psi^{'}(x_0) \sim c \land b \sim_{\phi} c \land b \sim_{\neg \textsc{B}}  c_1 \land \psi^{'}(b) \sim_{\phi} c) ~\lor \\ 
~  (x_0 \models \phi \land x_0 \models \textsc{B} \land b \sim_{\phi} c \land b \sim_{\neg \textsc{B}}  c_1 \land \psi^{'}(b) \sim_{\phi} c)  )
\end{array}$  \\
\hline

(17) $(\sim_{\phi} \in \{ >, \geq\} \land \sim_{\verb+B+}  \in \{  >, \geq \} \land \verb+s+_1 \uparrow \land \verb+s+_2  \downarrow)~~$ & \multirow{ 2}{*}{$(x_0 \models \phi \land x_0 \models \textsc{B}) $}   \\ (18) $(\sim_{\phi} \in \{ <, \leq\} \land \sim_{\verb+B+}  \in \{  <, \leq \} \land \verb+s+_1   \downarrow \land \verb+s+_2   \uparrow)~~$ & \\ 
\hline
(19) $(\sim_{\phi} \in \{ <, \leq\} \land \sim_{\verb+B+}  \in \{>, \geq\} \land \verb+s+_1   \uparrow \land \verb+s+_2 \downarrow)~~$ & \multirow{ 2}{*} {$(x_0 \models \phi \land x_0 \not\models \textsc{B}$)} \\ (20) $(\sim_{\phi} \in \{>, \geq\} \land \sim_{\verb+B+}  \in \{ <, \leq\} \land \verb+s+_1   \downarrow \land \verb+s+_2   \uparrow)~~$ & \\
\hline

(21) $(\sim_{\phi} \in \{ <, \leq\} \land \sim_{\verb+B+}  \in \{  <, \leq \} \land \verb+s+_1  \uparrow \land \verb+s+_2   \downarrow)~~$
& \multirow{ 2}{*} {Special decision procedure (see Algorithm \ref{alg:specialProcedure4.3.2})} \\ (22) $(\sim_{\phi} \in \{ <, \leq\} \land \sim_{\verb+B+}  \in \{>, \geq\} \land \verb+s+_1   \downarrow \land \verb+s+_2   \uparrow)~~$ & \\ 
\hline

(23) $(\sim_{\phi} \in \{>, \geq\} \land \sim_{\verb+B+}  \in \{ <, \leq\} \land \verb+s+_1 \uparrow \land \verb+s+_2  \downarrow$) ~~
& \multirow{ 2}{*} {Special decision procedure (see Algorithm \ref{alg:specialProcedure4.3.4})} \\ (24)  $(\sim_{\phi} \in \{ >, \geq\} \land \sim_{\verb+B+}  \in \{  >, \geq \} \land \verb+s+_1   \downarrow \land \verb+s+_2$   $\uparrow)$ ~~ & \\ 
\hline

~~(25-28) $( \sim_{\phi}, \sim_{\verb+B+}  \in \{<, \leq, >, \geq \} \land \verb+s+_1  \rightarrow_{b_1} \land~  \verb+s+_2  \rightarrow_{b_2})~$ & $\begin{array}[t]{l} 
 
 ( (x_0 \models \verb+B+ \land b_1 \sim_{\phi} c \land b_1 \sim_{\textsc{B}}  c_1) ~ \lor 
  (x_0 \not\models \verb+B+ \land b_2 \sim_{\phi} c \land b_2 \sim_{\neg \textsc{B}}  c_1) ~ \lor \\ 
  ~ (b_1 \sim_{\phi} c \land b_2 \sim_{\phi} c))  \end{array}$   \\
\hline

(29-32) $(\sim_{\phi}, \sim_{\verb+B+}  \in \{<, \leq, >, \geq \} \land \verb+s+_1  \uparrow \land~  \verb+s+_2  \uparrow) ~~~~~$  & See Observation \ref{Obs1} \\

(33-36) $(\sim_{\phi}, \sim_{\verb+B+}  \in \{<, \leq, >, \geq \} \land \verb+s+_1  \downarrow \land~  \verb+s+_2  \downarrow) ~~~~~$  &  \\ 
\hline 
\end{tabular} }
\caption{The termination rules of all possible cases for loop programs of the form (\ref{cyclesWithCond1})   \label{table:verification}}
\end{table}

\begin{example}
Consider the program:
$$
\textbf{while} (x \geq 5) ~ \{ \textbf{if}~ (x \geq 10)~ \textbf{then}~ x:= x+1; \textbf{else} ~ x:= x-1; \} 
$$
with $x_0 = 15$. Note that the syntactic structure of the given program matches with case 17 in Table \ref{table:verification} and hence we need to use the rule $(x_0 \models \phi \land x_0 \models \textsc{B}) $ to verify termination.  As one can see the rule holds for the given program and hence the loop is non-terminating.

\end{example}

\begin{example}
Consider the program:
$$
\textbf{while} (x \leq 10) ~ \{ \textbf{if}~ (x \leq 5)~ \textbf{then}~ x:= x+2; \textbf{else} ~ x:= x-3; \} 
$$
with $x_0 = 3$. For this program we will use Algorithm \ref{alg:specialProcedure4.3.2} to verify termination. Note that after five iterations (i.e. $x =3, 5, 7, 4, 6, 3$) the search will reach a fixed point with $x =3$ and hence the program is non-terminating. 
\end{example}

\section{Experimental Results} \label{sec:experimentalResults}

In this section, we give a brief description about the implementation of the approach, the tools used in the comparisons, and the conclusions drawn from the evaluation results obtained from the selected termination tools.

\subsection{The implementation of the proposed approach}

As mentioned earlier, the paper takes a different strategy to deal with common case of loop programs (monotone programs): rather than using ranking function based approach to decide termination, we apply check static analysis types of pattern matching.  Therefore, the termination problem for such class of programs is reduced to the satisfiability of a math-formula, i.e., a boolean combination of loop variables and linear mathematical relations over these variables.  The implementation of the approach consists of the following three phases.

\begin{itemize}

\item First, a simple program slicing technique is used to prune all statements in the analysed program that are not related to termination, leaving only those which can affect termination of the program.

\item Second, a static analysis types of pattern matching based on the techniques proposed in \cite{Spezialetti1995} is used to determine the class of loop monotonic variables (i.e. whether the loop counter is updated according to a monotonically increasing expression, monotonically decreasing expression, or a constant expression).

\item Then by a case distinction on the shape and the direction of the monotonic loop counter updates, termination is decided for given initial values using the termination rules derived in this paper (see Table \ref{table:verification}).

\end{itemize}


\subsection{Brief description about the tools used in the comparison}

We compare our tool with two state-of-the-art tools: (1) AProVE \cite{AProveIJCAR14} which is a system for automated termination and complexity proofs of term rewrite systems (TRSs), and (2) 2LS \cite{cdksw2015} a CPROVER-based framework, which reduces program analysis problems expressed in second order logic such as invariant or ranking function inference to synthesis problems over templates.  In the 5th Competition on Software Verification (SV-COMP'16), AProVE was the strongest tool for the termination category, while 2LS has been shown to be a powerful tool for proving termination for larger programs with thousands of lines of code \cite{cdksw2015}. We evaluate the tools using a large number of ANSI-C programs including:

\begin{itemize}

\item The SNU real-time benchmark suite that contains small C programs used for worst-case execution time analysis which are available at \path{www.cprover.org/goto-cc/examples/snu.html};

\item The Power-Stone benchmark suite as an example set of C programs for embedded systems \cite{Ku2007}.

\end{itemize}

During the evaluation we make the following manual simplifications: (1) we move the loops in the programs to the main function, (2) we simplify the complex statements and functions that do not affect the termination of the loop (e.g., make the function only contains one return statement), and (3) we add some variables declaration and initialization to ensure the correctness of the program. Our simplifications do not affect the outcome of termination of programs. 

\subsection{The evaluation results}

All experiments were run on Ubuntu 14.10, 3.2 GHZ Intel Core 4 duo CPU with 8GB RAM.  The results for SNU and Power-Stone are presented in Tables \ref{table:SNU} and \ref{table:PowerStone}. Each table reports the number of loops that were proven as terminating (T), non-terminating (NT), time-out (TO), or Maybe (M).  
Note that we set the timeout value to 300s. In most cases, the tools decide within a few seconds. 
As we observe in almost every program the proposed approach requires cheaper computational effort to verify termination. The verification time is reported in millisecond for our tool while it is reported in seconds for the other tools.   
It is interesting to mention that for the analysed programs of the SNU benchmark suite, AProVE could handle only 23 programs while 2LS could handle 45 programs. On the other hand, for the Power-Stone benchmark suite, AProVE could handle only 30 programs while 2LS could handle 53 programs.  More detailed information on the results and  performance of the tools is available at \path{ https://sites.google.com/site/termresult}.

Our implementation outperforms both AProVE and 2ls in two aspects. First, it can handle a larger
class of monotone programs than both AProVE and 2ls. Second, for the programs that all tools can handle our tool can decide termination much faster. The key advantage of the proposed approach is that it uses a light-weight static analysis technique based on recurrence equations to reason about termination, while the other tools use synthesis approaches based on ranking techniques.   Finding a ranking function of a loop program and proving its correctness is often expensive. The tools need to infer ranking functions and invariants on demand by iterating the loops. 
The analysis shows that tools based on ranking techniques  may fail to find a ranking function even for simple monotone loop programs as shown in Tables \ref{table:SNU} and \ref{table:PowerStone}.  

During the analysis, we note also that the tool AProVE does not handle very well programs with complex statements (i.e. those which contain pointers, arrays, and other data structures) even that the update statements that affect termination are simple monotone statements. For such cases the tool generates ``Maybe'' as an output. 
Finally, it is interesting to mention that the proposed static technique can be easily integrated into existing tools to handle common cases (monotone programs) in a much faster way.

\begin{table}  [h!]
\tiny
\adjustbox{max width=\textwidth}{
\begin{minipage}{0.60 \textwidth}
\begin{tabular}{|c|c|c|c|c|c|c|}
\hline
\textbf{Programs$\backslash$Tools} & \multicolumn{2}{c|}{Our tool}  & \multicolumn{2}{c|}{AProVE} &\multicolumn{2}{c|}{2ls} \\
\hline
 & result & time(\textbf{ms})  & result & time(s)  & result & time(s) \\
\hline
adpcm\_loop12.c & T & 1.173  & T & 6.18  & T & 0.76  \\
\hline
fft1\_loop6.c & T & 1.186 & M & 1.12 &  T & 0.40  \\
\hline
minver\_loop2.c & T & 1.601 & M & 1.10 &  T & 0.17  \\
\hline
crc\_loop3.c & T & 1.765 & T & 54.27  & M & 0.42  \\
\hline
minver\_loop6.c & T & 2.209 & M & 1.10 &  T & 0.23  \\
\hline
fft1\_loop1.c & T & 0.904 & M & 1.18 &  T & 0.18  \\
\hline
fft1\_loop4.c & T & 0.915 & T & 2.41  & T & 0.18  \\
\hline
minver\_loop7.c & T & 1.795 & M & 1.16 &  T & 0.20  \\
\hline
fft1\_loop5.c & T & 1.493 & M & 1.14 &  TO & 300.06 \\
\hline
adpcm\_loop11.c & T & 0.955 & T & 2.69  & T & 0.48  \\
\hline
adpcm\_loop9.c & T & 1.306 & T & 2.19  & T & 0.17  \\
\hline
fft1\_loop2.c & T & 0.848 & M & 1.34 &  T & 0.19  \\
\hline
fft1\_loop3.c & T & 1.121 & M & 1.07 &  M & 0.77 \\
\hline
matmul\_loop1.c & T & 1.375 & M & 2.89 &  T & 0.22  \\
\hline
crc\_loop1.c & T & 1.318 & T & 10.28  & T & 0.16  \\
\hline
adpcm\_loop5.c & T & 1.091 & T & 9.17  & T & 0.56  \\
\hline
minver\_loop1.c & T & 1.666 & M & 1.18 &  T & 0.15  \\
\hline
fft1k\_loop4.c & T & 1.807 & M & 1.04 &  M & 9.01  \\
\hline
adpcm\_loop17.c & T & 1.342 & M & 1.17 &  T & 1.27  \\
\hline
minver\_loop9.c & T & 2.118 & M & 1.19 &  T & 0.53  \\
\hline
lms\_loop6.c & T & 1.137 & M & 1.12 &  M & 0.36  \\
\hline
adpcm\_loop16.c & T & 1.286 & M & 1.23 &  T & 1.15  \\
\hline
ludcmp\_loop4.c & T & 8.527 & M & 1.32 &  T & 0.41  \\
\hline
lms\_loop7.c & T & 1.093 & M & 1.24 &  T & 0.18  \\
\hline
ludcmp\_loop5.c & T & 1.178 & M & 1.10 &  TO & 300.06  \\
\hline
fft1k\_loop2.c & T & 0.848 & T & 1.98  & T & 0.18  \\
\hline
adpcm\_loop4.c & T & 0.978 & M & 2.01 &  T & 0.12  \\
\hline
adpcm\_loop1.c & T & 1.003 & T & 2.07  & T & 0.11  \\
\hline
ludcmp\_loop1.c & T & 1.568 & M & 1.29 &  TO & 300.02 \\
\hline
fir\_loop1.c & T & 0.928 & T & 1.97  & T & 0.13  \\
\hline
minver\_loop3.c & T & 2.326 & M & 1.10 &  T & 0.43  \\
\hline
adpcm\_loop10.c & T & 1.630 & T & 2.80  & T & 0.14  \\
\hline
adpcm\_loop15.c & T & 1.364 & M & 1.23 &  M & 3.44  \\
\hline
adpcm\_loop2.c & T & 1.209 & T & 1.98  & T & 0.14  \\
\hline
adpcm\_loop14.c & T & 1.229 & T & 22.24  & T & 0.24  \\
\hline
adpcm\_loop3.c & T & 1.241 & T & 8.87  & T & 0.63  \\
\hline
qsort-exam\_loop1.c & T & 1.882 & M & 1.13 &  M & 0.28  \\
\hline
minver\_loop4.c & T & 2.138 & M & 1.14 &  T & 0.23  \\
\hline
adpcm\_loop13.c & T & 1.032 & T & 2.51  & T & 0.16  \\
\hline
minver\_loop5.c & T & 1.169 & M & 2.27 &  T & 0.39  \\
\hline
minver\_loop8.c & T & 1.781 & M & 1.11 &  T & 0.27  \\
\hline
adpcm\_loop7.c & T & 1.133 & T & 5.29  & T & 0.13  \\
\hline
fir\_loop5.c & T & 1.195 & M & 1.06 &  M & 0.82  \\
\hline
matmul\_loop2.c & T & 1.412 & M & 8.78 &  T & 0.30  \\
\hline
fir\_loop2.c & T & 0.899 & T & 2.13  & T & 0.18  \\
\hline
lms\_loop4.c & T & 1.330 & M & 1.14 &  TO & 300.22 \\
\hline
lms\_loop5.c & T & 1.146 & M & 1.15 &  TO & 300.22  \\
\hline
adpcm\_loop8.c & T & 1.374 & T & 6.31  & T & 0.13  \\
\hline
fft1k\_loop1.c & T & 0.916 & T & 1.95  & T & 0.21  \\
\hline
jfdctint\_loop1.c & T & 0.981& T & 2.47  & T & 0.39  \\
\hline
adpcm\_loop6.c & T & 1.024 & M & 2.41 &  T & 0.17  \\
\hline
fft1k\_loop3.c & T & 2.066 & M & 1.21 &  TO & 300.02  \\
\hline
fir\_loop4.c & T & 1.355 & M & 1.15 &  M & 1.08  \\
\hline
ludcmp\_loop2.c & T & 1.194 & M & 1.28 &  T & 3.73  \\
\hline
fir\_loop3.c & T & 1.046 & M & 1.17 &  TO & 300.22  \\
\hline
ludcmp\_loop3.c & T & 1.177 & M & 1.15 &  TO & 300.04  \\
\hline
jfdctint\_loop2.c & T & 1.919 & TO & 300.06 &  T & 0.83  \\
\hline
jfdctint\_loop3.c & T & 9.560 & TO & 300.05 &  T & 0.74  \\
\hline
lms\_loop2.c & T & 0.898 & T & 2.14  & T & 0.17  \\
\hline
lms\_loop1.c & T & 1.074 & T & 1.98  & T & 0.13  \\
\hline
fibcall\_loop1.c & T & 0.995 & T & 44.49  & M & 6.24  \\
\hline
lms\_loop3.c & T & 1.261 & M & 1.10 &  M & 0.85  \\
\hline
crc\_loop2.c & T & 1.612 & TO & 300.05 &  T & 0.33  \\
\hline
Total & 63 & 99.102 & 23 & 1154.1 & 45 & 2442.33 \\
\hline

\end{tabular}

\captionsetup{font=tiny}
\caption{SNU real-time benchmark suite}
\label{table:SNU}

\end{minipage}

\begin{minipage}{0.55 \textwidth}

\begin{tabular}{|c|c|c|c|c|c|c|}
\hline
\textbf{Programs$\backslash$Tools} & \multicolumn{2}{c|}{Our tool}  & \multicolumn{2}{c|}{AProVE} &\multicolumn{2}{c|}{2ls} \\
\hline
 & result & time(\textbf{ms}) & result & time(s) & result & time(s) \\
\hline
adpcm\_loop12.c & T & 2.494  & M & 13.64  & T & 0.29  \\
\hline
blit\_loop3.c & T & 1.244  & M & 2.01  & T & 0.14  \\
\hline
compress\_loop6.c & T & 1.312  & TO & 319.34  & T & 0.22  \\
\hline
huff\_loop1.c & T & 1.221  & T & 4.85  & T & 0.19  \\
\hline
compress\_loop7.c & T & 1.569  & M & 8.84  & T & 26.41  \\
\hline
compress\_loop8.c & T & 1.117  & M & 1.68  & T & 18.54  \\
\hline
pocsag\_loop2.c & T & 2.110  & M & 1.13  & T & 0.66  \\
\hline
adpcm\_loop11.c & T & 1.787  & TO & 306.37  & T & 0.93  \\
\hline
huff\_loop2.c & T & 1.500  & T & 16.92  & T & 1.46  \\
\hline
blit\_loop1.c & T & 1.188  & M & 1.25  & T & 0.49  \\
\hline
adpcm\_loop9.c & T & 1.695  & T & 2.57  & T & 0.48  \\
\hline
huff\_loop3.c & T & 1.375  & M & 4.36  & T & 6.30  \\
\hline
g3fax\_loop1.c & T & 1.108  & T & 6.53  & T & 1.50  \\
\hline
compress\_loop4.c & T & 1.120  & T & 2.89  & T & 22.48  \\
\hline
compress\_loop1.c & T & 1.149  & T & 2.38  & T & 1.04  \\
\hline
jpeg\_loop10.c & T & 1.056  & T & 2.47  & T & 0.18  \\
\hline
crc\_loop1.c & T & 1.352  & T & 9.84  & T & 0.17  \\
\hline
compress\_loop5.c & T & 1.101  & T & 2.02  & T & 12.30  \\
\hline
huff\_loop4.c & T & 1.718  & T & 4.51  & T & 2.77  \\
\hline
compress\_loop2.c & T & 1.034  & M & 2.93  & T & 0.58  \\
\hline
compress\_loop3.c & T & 1.045  & M & 6.07  & T & 0.88  \\
\hline
ucbqsort\_loop4.c & T & 0.960  & M & 1.43  & M & 0.13  \\
\hline
adpcm\_loop4.c & T & 1.938  & M & 2.18  & T & 0.13  \\
\hline
adpcm\_loop1.c & T & 1.622  & T & 10.14  & T & 0.25  \\
\hline
adpcm\_loop10.c & T & 1.873  & M & 1.14  & T & 2.15  \\
\hline
adpcm\_loop5.c & T & 1.593  & T & 5.50  & T & 0.15  \\
\hline
ucbqsort\_loop3.c & T & 0.981  & M & 1.24  & M & 0.15  \\
\hline
adpcm\_loop2.c & T & 1.480  & M & 1.82  & T & 0.18  \\
\hline
ucbqsort\_loop2.c & T & 1.056  & M & 1.39  & M & 0.13  \\
\hline
adpcm\_loop3.c & T & 1.468  & T & 22.39  & T & 0.29  \\
\hline
v42\_loop5.c & T & 1.515  & M & 1.06  & TO & 300.71  \\
\hline
ucbqsort\_loop1.c & T & 1.420  & T & 6.51  & T & 0.56  \\
\hline
v42\_loop3.c & T & 0.941  & T & 1.98  & T & 0.19  \\
\hline
jpeg\_loop3.c & T & 1.343  & M & 2.20  & T & 0.12  \\
\hline
pocsag\_loop1.c & T & 1.075  & T & 1.83  & T & 0.15  \\
\hline
pocsag\_loop4.c & T & 1.479  & T & 2.43  & T & 0.57  \\
\hline
jpeg\_loop2.c & T & 1.265  & T & 9.07  & T & 0.21  \\
\hline
jpeg\_loop7.c & T & 1.277  & M & 2.23  & T & 0.13  \\
\hline
pocsag\_loop5.c & T & 1.788  & T & 1.49  & T & 0.23  \\
\hline
jpeg\_loop6.c & T & 1.110  & T & 2.08  & T & 0.19  \\
\hline
v42\_loop2.c & T & 0.952  & T & 1.75  & T & 0.17  \\
\hline
fir\_loop2.c & T & 1.758  & M & 1.24  & M & 1.68  \\
\hline
adpcm\_loop7.c & T & 1.404  & T & 2.21  & T & 0.17  \\
\hline
jpeg\_loop1.c & T & 1.163  & T & 2.60  & T & 0.24  \\
\hline
adpcm\_loop8.c & T & 1.439  & T & 2.85  & T & 0.12  \\
\hline
v42\_loop4.c & T & 1.518  & M & 1.20  & TO & 300.35  \\
\hline
adpcm\_loop6.c & T & 1.518  & T & 4.96  & T & 0.15  \\
\hline
fir\_loop1.c & T & 1.148  & M & 1.24  & TO & 300.31  \\
\hline
pocsag\_loop3.c & T & 0.974  & T & 1.81  & T & 0.11  \\
\hline
jpeg\_loop11.c & T & 1.033  & T & 7.39  & T & 10.35  \\
\hline
fir\_loop3.c & T & 1.191  & M & 1.15  & M & 1.38  \\
\hline
blit\_loop2.c & T & 1.450  & M & 1.26  & T & 0.37  \\
\hline
jpeg\_loop8.c & T & 2.366  & T & 11.33  & T & 21.05  \\
\hline
huff\_loop5.c & T & 1.836  & M & 9.61  & T & 1.86  \\
\hline
pocsag\_loop6.c & T & 1.535  & M & 5.07  & T & 0.41  \\
\hline
v42\_loop6.c & T & 1.440  & M & 1.94  & M & 1.39  \\
\hline
pocsag\_loop8.c & T & 2.197  & M & 1.17  & T & 0.45  \\
\hline
v42\_loop1.c & T & 1.328  & T & 18.66  & T & 0.15  \\
\hline
jpeg\_loop5.c & T & 1.148  & T & 6.28  & T & 0.14  \\
\hline
jpeg\_loop4.c & T & 1.363  & M & 2.66  & T & 0.20  \\
\hline
jpeg\_loop9.c & T & 1.182  & T & 2.57  & T & 5.45  \\
\hline
pocsag\_loop7.c & T & 1.246  & M & 7.05  & T & 0.27  \\
\hline
crc\_loop2.c & T & 1.576  & T & 61.93  & M & 0.58  \\
\hline
Total & 63 & 88.244 & 30 & 958.64 & 53 & 1051.98  \\
\hline

\end{tabular}

\captionsetup{font=tiny}
\caption{PowerStone benchmark suite} 
\label{table:PowerStone}

\end{minipage}

}

\end{table}

\section{Conclusion and Future Work} 

We presented an efficient approach to prove termination of monotone programs, a class of loops
that is often encountered in computer programs.    The approach uses a light-weight static analysis technique based on recurrence equations  and takes advantage of properties of monotone functions.
The proposed approach leads to significant performance improvement against previous tools where they are applicable.
In future work, we aim to extend the tool to support a richer class of monotone programs, in particular we aim to study termination of multi-path programs that are  more general than the ones considered in this paper, but for which termination is still decidable.  Furthermore, since the termination problem is obviously related to the worst-case execution time problem (WCET): a loop program that does not terminate has an unbounded WCET, we aim also to extend the tool to reason about WCET of monotone loops while taking advantage of the termination rules derived in the paper.

\bibliographystyle{plain}
\bibliography{references}

\end{document}